\documentclass[showpacs,preprintnumbers,amsmath,amssymb,preprint]{revtex4}

\usepackage{graphicx}
\usepackage{color}
\usepackage{amsmath,amssymb,amssymb}

\begin{document}

\title{Identifying topological-band insulator transitions in silicene and other 2D gapped Dirac materials  by means of  
R\'enyi-Wehrl entropy }

\author{M. Calixto}
\affiliation{Departamento de Matem\'atica Aplicada, Universidad de Granada,
Fuentenueva s/n, 18071 Granada, Spain}
\author{E. Romera}
\affiliation{Departamento de F\'{\i}sica At\'omica, Molecular y Nuclear and
Instituto Carlos I de F{\'\i}sica Te\'orica y
Computacional, Universidad de Granada, Fuentenueva s/n, 18071 Granada,
Spain}


\begin{abstract}

We propose a new method to identify  transitions from a topological insulator to a band insulator in silicene
(the silicon equivalent of graphene) in the presence of  perpendicular magnetic and electric fields, 
by using the R\'enyi-Wehrl entropy of the quantum state in phase space. Electron-hole entropies  
display an inversion/crossing behavior at the charge neutrality point for any Landau level, and  
the combined entropy of particles plus holes turns out to be maximum at this critical point. The result 
is interpreted in terms of delocalization of the quantum state in phase space. The entropic description 
presented in this work  will be valid in general 2D gapped Dirac materials, 
with a strong intrinsic spin-orbit interaction, isoestructural with silicene.
\end{abstract}

\maketitle

\section{Introduction}

Silicene is a two dimensional crystal of silicon which has been
studied theoretically \cite{takeda,guzman94} and recently experimentally
 \cite{vogt12,Aufray10,lalmi10,fleurence12,padova}. The low energy
electronic properties can be described by a Dirac Hamiltonian, like in graphene, but the
electrons are massive  due to a relative
large spin-orbit coupling $\Delta_\mathrm{so}$. In fact, silicene has a band gap $|\Delta_{s\xi}|$ ($s$ and $\xi$ denote spin and valley, respectively) which can be
controlled by applying a perpendicular electric field $\mathcal{E}_z=\Delta_z/l$ ($l$ is the interlattice distance of the buckled honeycomb structure) to the silicene sheet 
(see  \ref{gapsili} for a plot of the band gap). 
It has been demonstrated \cite{tahir2013} that there is a  quantum phase transition, from a topological  
insulator (TI, $|\Delta_z|<\Delta_\mathrm{so}$) to a band insulator (BI, $|\Delta_z|>\Delta_\mathrm{so}$), at a charge neutrality
point (CNP)  $\Delta_z^{(0)}=s\xi\Delta_\mathrm{so}$, where there is a gap cancellation  between the perpendicular electric field
and the spin-orbit coupling, thus exhibiting a semimetal behavior. 

A 2D topological insulator is also known as a quantum spin Hall state, and it was theoretically studied in \cite{Kane} and first 
discovered in HgTe quantum wells in \cite{Bernevig}. The common characteristic of a TI-BI transition is a band inversion with a level crossing at some 
critical value of a control parameter (electric field, quantum well thickness, etc). In this paper we find that electron-hole entropies 
of a quantum state (in a phase-space representation) also exhibit this crossing/inversion behavior at the CNP, thus 
providing a new characterization of the TI-BI quantum phase transition (QPT).

The study of phase space properties in quantum systems can
be done by using the Husimi function of a quantum state $|\psi\rangle$, which is defined  as the squared overlap between a
minimal uncertainty (coherent) state $|\alpha\rangle$ and $|\psi\rangle$, thus giving the
probability of finding  the quantum system in a coherent state.  
Husimi quasiprobability distribution has been useful  for a phase-space analysis
of metal-insulator transition \cite{aulbach},  for the study of quantum chaos in
physics \cite{monteiro}, and
recently in the analysis of QPTs in algebraic models \cite{husidi,husivi,LMG,EPL}. 
R\'enyi-Wehrl entropies of the  Husimi function
  (see section \ref{renyiwehrl} for a definition) have been proved to be a good marker for QPTs of some 
paradigmatic models like: Dicke model  \cite{husidi} of atom-field interactions 
(undergoing a QPT from normal to superradiant), vibron model \cite{husivi} for triatomic molecules 
(undergoing a shape QPT from linear to bent)  and 
Lipkin-Meshkov-Glick model \cite{LMG} used for example  in nuclear
physics and quantum optics. R\'enyi-Wehrl entropy displays either 
an extreme, or an abrupt change, at the critical point and therefore provides a detector of QPTs  
across the phase diagram of a critical system. 

In this work we shall analyze the topological-band insulator transition in 
silicene by using these phase-space techniques. The paper is organized as
follows. Firstly, in Section \ref{Hamilsec}, we shall remind the low energy Hamiltonian for
silicene, its eigenvectors and eigenvalues. Then,  in Section \ref{entropysec}, we will define the Husimi function 
in this model and the  R\'enyi-Wehrl entropies of the Husimi function. In Section \ref{marker} we will show 
that the R\'enyi-Wehrl entropy turns out  to be a clear marker of
topological-band insulator transitions in silicene. Finally, some concluding
remarks will be given in the last Section.

\section{Low energy Hamiltonian}\label{Hamilsec}

\begin{figure}
\includegraphics[width=8cm]{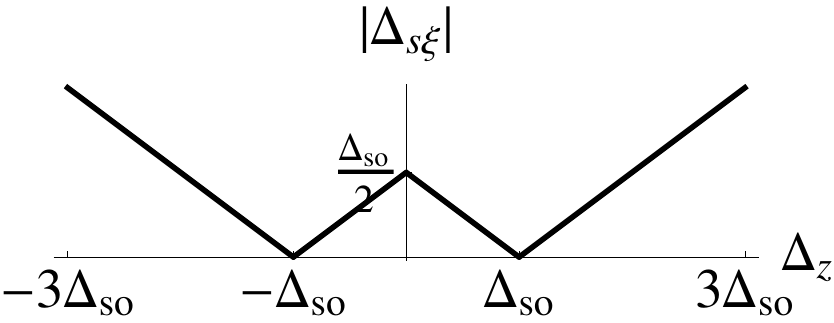}
\caption{ \label{gapsili} The band gap $\Delta_{s\xi}=(\Delta_z-s\xi\Delta_{\mathrm{so}})/2$ as a function of the electric potential $\Delta_z$ 
created by the perpendicular electric field applied to the silicene sheet.  
$\Delta_\mathrm{so}$ is the spin-orbit energy gap induced by the intrinsic spin-orbit.}
\end{figure}

Let us consider a monolayer silicene with  external magnetic    $B$ and electric
 $\mathcal{E}_z$ fields  applied perpendicular to the  sheet. The effective low energy Hamiltonian is given by \cite{tahir2013}
\begin{equation}
H_s^{\xi}=v_F (\sigma_x  p_x-\xi \sigma_y  p_y )-\xi
s \Delta_{\mathrm{so}} \sigma_z+ \Delta_z \sigma_z,
\label{hamiltoniano}
\end{equation}
where $\xi$ corresponds  to the inequivalent corners ${K}$ ($\xi=1$) and
${K}^{\prime}$  ($\xi=-1$) of the Brillouin zone, respectively, 
${\sigma}_j$ are the usual Pauli matrices, $v_F=5\times 10^5$ m/s is the Fermi
velocity of the Dirac fermions  (with up and down spin values being
represented by $s=\pm 1$, respectively) and  $\Delta_{\rm so}$ is the band gap  induced by
intrinsic spin-orbit interaction, which provides a mass to the Dirac fermions. The spin-orbit energy gap induced by intrinsic  
spin-orbit coupling  has been  estimated  (using density functional theory  and tight-binding calculations 
\cite{drummond2012,liu2011,liub2011})  as $\Delta_{\mathrm{so}}\approx
1.55-7.9$ meV, and  we will consider $\Delta_{\mathrm{so}}=4$ meV along this paper.  
The application of an external constant electric field $\mathcal{E}_z$ creates an potential difference $\Delta_z=l\mathcal{E}_z$, 
with $l\simeq 0.23\AA$, between sublattices yielding a tunable band gap. See  Figure \ref{gapsili} for a graphical representation. 

Using the Landau gauge, $\vec{A}=(0, Bx,0)$,  and doing the minimal coupling $\vec{p}\to \vec{p}+\frac{e}{c}\vec{A}$ for the momentum, 
one can easily diagonalize  the Hamiltonian (\ref{hamiltoniano}) with eigenvalues 
 \cite{tahir2013,stille2012}
\begin{equation}
E_{n}^{s\xi}=\left\{\begin{array}{l} \mathrm{sgn}(n) \sqrt{|n|\hbar^2\omega^2 + \Delta_{s\xi}^2}, \quad n\neq 0, \\ 
              -\xi\Delta_{s\xi}, \quad n= 0, \end{array}\right.\label{especteq}
            \end{equation}
in terms of the Landau level index $n=0,\pm 1,\pm 2,\dots$, the cyclotron frequency 
$\omega=v_F\sqrt{2eB/\hbar}$  and the lowest band gap $\Delta_{s\xi}\equiv(\Delta_z-s\xi\Delta_{\mathrm{so}})/2$. 
The corresponding 
eigenvectors for the $K$  and $K'$ points are 
\begin{equation}
|n\rangle_{s\xi}=\left(\begin{array}{c}
-i A_{n}^{s\xi}||n|-(1+\xi)/2\rangle\\
B_{n}^{s\xi}||n|-(1-\xi)/2\rangle
\end{array}\right),
\label{vectors}
\end{equation}
where the constants $A_{n}^{s\xi}$ and $ B_{n}^{s\xi}$ are given by \cite{stille2012}
\begin{eqnarray}
 A_{n}^{s\xi}&=&\left\{\begin{array}{l}
 \mathrm{sgn}(n)\sqrt{\frac{|E_{n}^{s\xi}|+\mathrm{sgn}(n)\Delta_{s\xi}}{{2|E_{n}^{s\xi}|}}},  \quad n\neq 0, \\
(1-\xi)/2,  \quad n=0, 
\end{array}\right.\nonumber\\ 
 B_{n}^{s\xi}&=&\left\{\begin{array}{l}
\sqrt{\frac{|E_{n}^{s\xi}|-\mathrm{sgn}(n)\Delta_{s\xi}}{{2|E_{n}^{s\xi}|}}},  \quad n\neq 0, \\
(1+\xi)/2,  \quad n=0, 
\end{array}\right.\label{coef}
\end{eqnarray}
and $||n|\rangle$ denotes an orthonormal Fock state of the harmonic
oscillator.

\begin{figure}
\includegraphics[width=8cm]{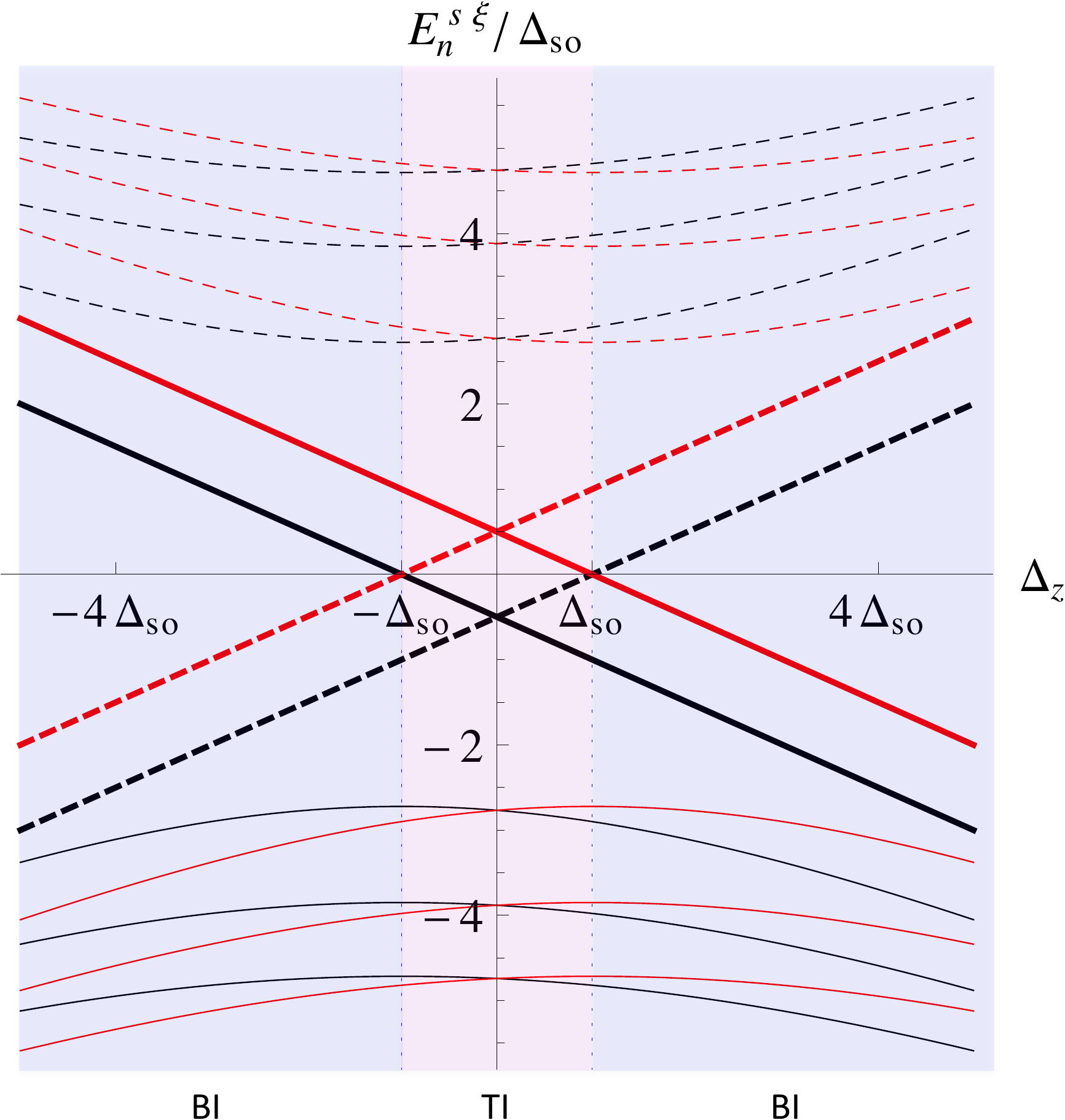}
\caption{Low energy spectra of silicene as a function of the external electric potential $\Delta_z$ for $B=0.5$ T. 
Landau levels $n=\pm 1, \pm 2$ and $\pm 3$, at valley $\xi=1$,  are represented by 
dashed (electrons) and solid (holes) thin lines, black for $s=-1$ and red for
$s=1$ (for the other valley we simply have $E_{n}^{s,-\xi}=E_{n}^{-s,\xi}$). The lowest Landau level $n=0$ is represented by thick lines at both valleys:  solid at $\xi=1$ and dashed 
at $\xi=-1$. Vertical blue dotted grid lines indicate the CNPs separating BI (blue) from TI (pink) phases.}
\label{energias}
\end{figure}

In Figure \ref{energias} we represent the low energy spectra given by eq. \eqref{especteq}, 
as a function of the external electric potential $\Delta_z$ for $B=0.5$ T (where T is the unit symbol for Tesla).
The transitions  TI-BI occur at
the critical points $\Delta_z=s\xi \Delta_{\mathrm{so}}$, where there is a band inversion 
for the $n=0$ Landau level (either for spin up and down) at both valleys. The
energies $E_0^{1,\xi}$ and $E_0^{-1,\xi}$ have the same sign in the BI phase
($|\Delta_z|>\Delta_\mathrm{so}$, in blue), and different 
sign in the TI phase ( $|\Delta_z|<\Delta_\mathrm{so}$, in pink).

\section{Inverse participation ratio and R\'enyi-Wehrl entropy \label{renyiwehrl}}\label{entropysec}

The standard canonical (or Glauber) coherent states for the 1D quantum harmonic oscillator  are given by
\begin{equation}
|\alpha\rangle=e^{-|\alpha|^2/2}e^{\alpha a^\dag}|0\rangle=
e^{-|\alpha|^2/2}\sum_{n=0}^\infty\frac{\alpha^n}{\sqrt{n!}}|n\rangle,
\end{equation}
with $\alpha\in\mathbb C$ (the phase space). The coherent states form an overcomplete set of the corresponding 
Hilbert space and verify the closure relation:
\begin{equation}
1=\int_{{\mathbb R}^2}|\alpha\rangle\langle\alpha|\frac{d^2\alpha}{\pi},\label{closure}
\end{equation}
with $d^2\alpha=d\mathrm{Re}(\alpha)d\mathrm{Im}(\alpha)$. Using the coherent state (Bargmann) representation, we can associate 
a quasiprobability distribution (the so called Husimi function) $Q_\psi(\alpha)\equiv|\langle\alpha|\psi\rangle|^2$ to every 
normalized harmonic oscillator state $|\psi\rangle$. This definition is straightforwardly extended to 
any density matrix $\rho$ as $Q_\rho(\alpha)=\langle\alpha|\rho|\alpha\rangle$. 
For example, the Husimi function of a Fock state $|n\rangle$ is simply 
\begin{equation}
 Q_{n}(\alpha)=|\langle n|\alpha\rangle|^2=\frac{e^{-|\alpha|^2}}{n!}|\alpha|^{2n}.
\label{husfock}
\end{equation}
Coherent states are said to be ``quasi-classical'' because of their minimum uncertainty (and dynamical) properties. 
To quantify the spread/localization of a wave packet $\psi$ in phase space, the $\nu$-th moments of the corresponding Husimi 
function $Q_\psi$, defined by
\begin{equation}
 M^{\nu}_\psi=\int_{{\mathbb R}^2}\frac{d^2\alpha }{\pi} (Q_\psi(\alpha))^{\nu},\label{momentsnu}
\end{equation}
are often used. Note that, using \eqref{closure}, we have $M^1_\psi=1$ for normalized $\psi$. 
We shall focus on the $\nu=2$ moment, also called the inverse participation ratio, 
which is related to the inverse area in phase space occupied by the Husimi function. Namely, for a Fock state $|n\rangle$ we have
\begin{equation}
 M^{2}_n=\frac{(2n)!}{2^{2n+1}(n!)^2},\label{moment2n}
\end{equation}
which is maximum for $n=0$, $M^2_0=1/2$ (minimum area). Instead of the momentum $M^{\nu}_\psi$, we shall use the 
R\'enyi-Wehrl entropy $W^{\nu}_\psi=\frac{1}{1-\nu}\ln(M^{\nu}_\psi)$ for the sake of convenience. Note that, for 
$\nu\to 1$, the previous expression reads $W^1_\psi=-\int
Q_\psi(\alpha)\ln\left( Q_\psi(\alpha)\right)d^2\alpha/{\pi}$, the so called 
Wehrl entropy. As conjectured by Wehrl \cite{Wehrl} and proved by Lieb \cite{Lieb}, any Glauber coherent
state $|\alpha'\rangle$ has a minimum Wehrl entropy of $W^1_{\alpha'}=1$. We shall use $\nu=2$, for which the minimum 
R\'enyi-Wehrl entropy is $W^2_{\mathrm{min}}=\ln(2)$, corresponding to a coherent state.

\section{R\'enyi-Wehrl entropy as a marker of topological-band
  insulator transitions \label{marker}}

\begin{figure}
\includegraphics[width=8cm]{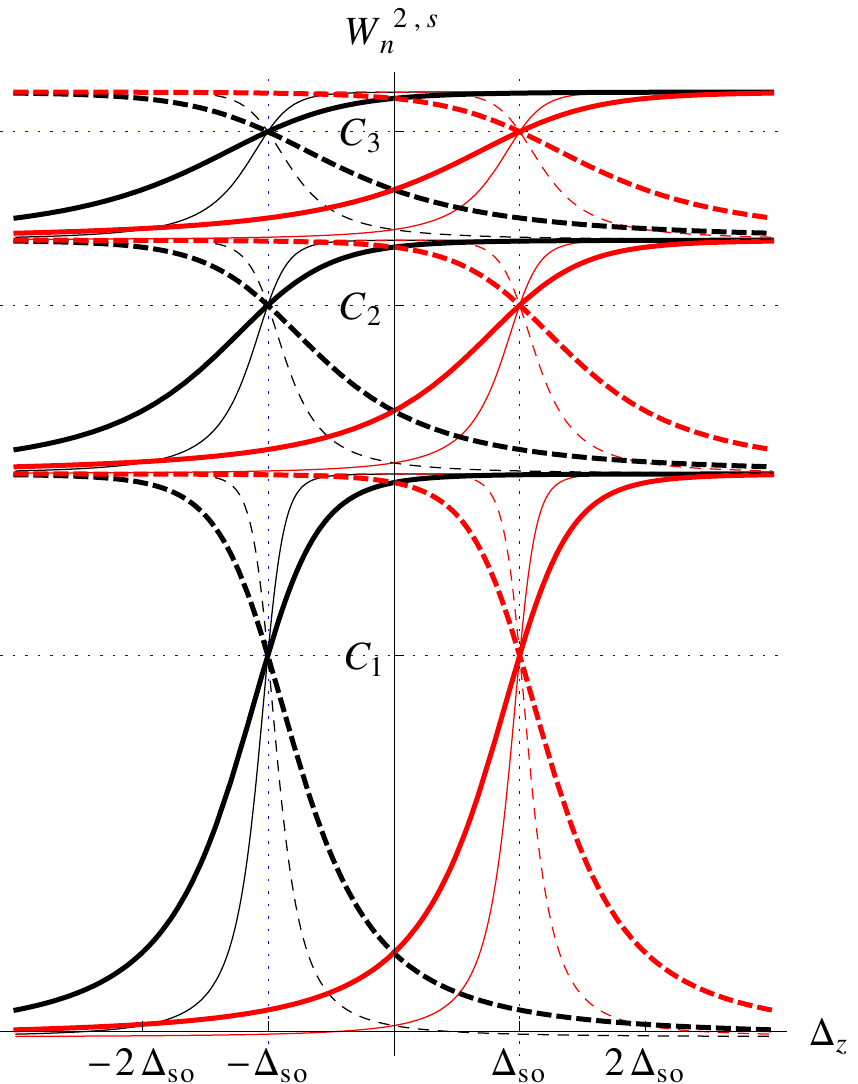}
\caption{ \label{Wcortes} R\'enyi-Wehrl entropy $W^{2,s}_n$ of a Hamiltonian eigenstate $|n\rangle_{s\xi}$ 
as a function of the electric potential $\Delta_z$ for the Landau levels: $n=1, 2$ and $3$  
(electrons; dashed curves) and $n=-1, -2$ and $-3$ 
(holes; solid curves) and valley $\xi=1$. Black and red curves correspond to spin $s=-1$ and $s=1$, respectively. Electron and hole entropy curves 
cross (``entropy inversion'') at the critical value of the electric potential $\Delta_z^{(0)}=-s\Delta_{\mathrm{so}}$ (vertical blue dotted grid lines indicate this CNPs), 
with crossing entropy values denoted by $C_n=W^{2,s}_n(\Delta_z^{(0)})$, $n=1,2,3$ and given in equation \eqref{maxent}
(horizontal blue dotted grid lines indicate these crossing points). Thin and thick lines 
correspond to magnetic fields $B=10^{-3}$ and $B=10^{-2}$ T, respectively.}
\end{figure}
Now, we shall calculate the R\'enyi-Wehrl entropy $W^2_n$ of a Hamiltonian eigenstate \eqref{vectors} as a function 
of the electric potential $\Delta_z$ in order to analyze the TI-BI transition. We shall restrict ourselves to the valley $\xi=1$, omitting this index  from 
\eqref{vectors} and \eqref{coef}. All the results for $\xi=1$ are straightforwardly translated to $\xi=-1$ by swapping 
electrons for holes (i.e., $n\leftrightarrow -n$) and spin up for down (i.e., $s\leftrightarrow -s$). 
Taking into account equations (\ref{vectors}) and (\ref{husfock}), we can write the expression of the Husimi function
of a Hamiltonian eigenstate  $|n\rangle_{s}$  as
\begin{equation}
Q_{n}^{s}(\alpha)=|\langle
\alpha|{n}\rangle_{s}|^2=(A_{n}^{s})^2 Q_{|n|-1}(\alpha)+(B_{n}^{s})^2 Q_{|n|}(\alpha),
\end{equation}
and the second moment as 
\begin{eqnarray}
M^{2,s}_n&=&(A_{n}^{s})^4 M^2_{|n|-1}+(B_{n}^{s})^2
M^2_{|n|}\nonumber\\ &&+2(A_{n}^{s}B_{n}^{s})^2 
 \frac{(2|n|-1)!}{4^{|n|}|n|!(|n|-1)!}
\end{eqnarray}
where we are using \eqref{moment2n} and the value of $\int\frac{d^2\alpha}{\pi} Q_{|n|}(\alpha)Q_{|n|-1}(\alpha)$. For the sake of convenience, 
we shall use the R\'enyi-Wehrl entropy $W^{2,s}_n=-\ln(M^{2,s}_n)$ to make graphical representations as a function of $\Delta_z$ for 
different values of the Landau level $n$, spin $s=\pm 1$ and the magnetic field $B$. In Figure \ref{Wcortes} we plot 
$W^{2,s}_n$ as a function of $\Delta_z$ for $n=1, 2, 3$ (electrons) and $n=-1, -2, -3$ (holes) with spin up $s=1$ (in red) and down $s=-1$ (in black). 
\begin{figure}
a)\includegraphics[width=8cm]{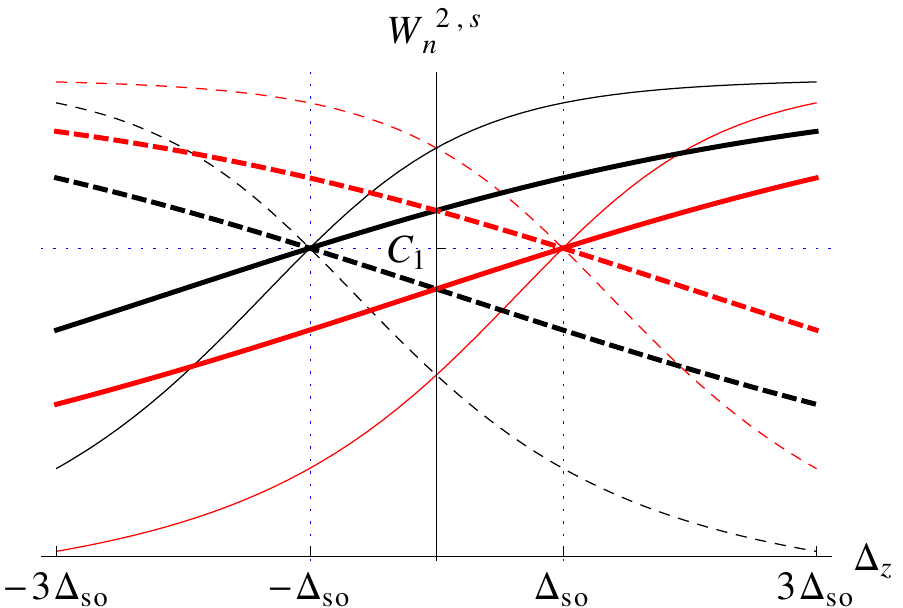}
b)\includegraphics[width=8cm]{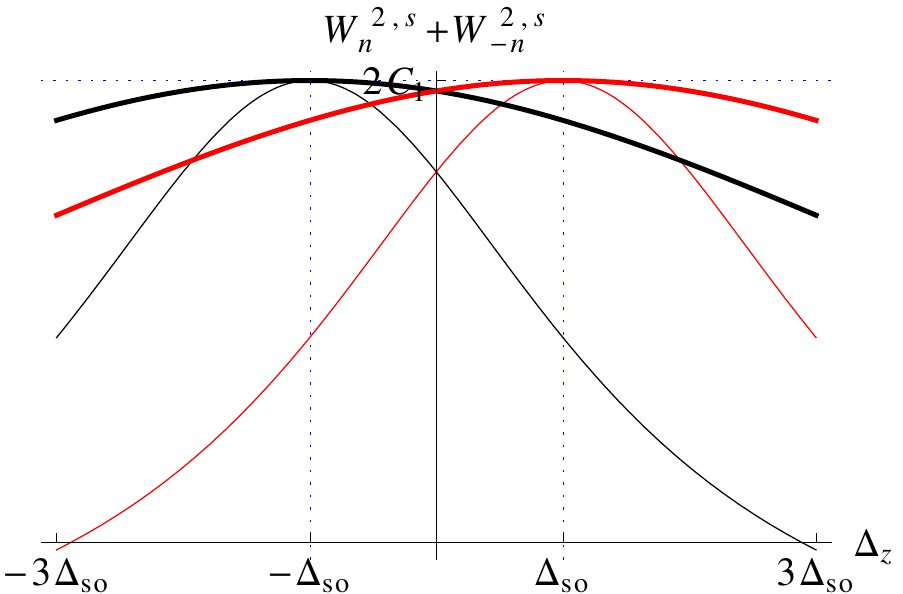}
\caption{ \label{Renyi1} a) Same representation as in Figure
  \ref{Wcortes} but for $n=\pm 1$ only and  magnetic fields $B=0.1$ 
  (thin) and $B=1$ T (thick), respectively.    
b) Combined entropy $W^{2,s}_n+W^{2,s}_{-n}$ of electrons plus holes for $n=1$. The entropy of the electron-hole compound 
exhibits a maximum of $2C_n$ at the CNP $\Delta_z^{(0)}=-s\Delta_{\mathrm{so}}$.}
\end{figure}
Entropy for holes ($n<0$, solid curves) is an increasing function of the electric potential $\Delta_z$, whereas entropy for electrons ($n>0$, dashed curves) is a decreasing function of $\Delta_z$. 
Electron-hole entropy curves cross at the CNP $\Delta_z^{(0)}=-s\Delta_{\mathrm{so}}$, where the entropy 
is
\begin{equation}
C_n\equiv W^{2,s}_{n}(\Delta_z^{(0)})=-\ln\left(\frac{(8|n|-3)\Gamma(|n|-\frac{1}{2})}{16\sqrt{\pi}\Gamma(|n|+1)}\right).\label{maxent}
\end{equation}
Note that the critical entropy \eqref{maxent} 
only depends on the Landau level $n$ (and not on other physical magnitudes like: $B, v_F, e, s$, etc) thus providing a universal characterization of the 
topological-band insulator QPT, which is shared with other two-dimensional crystals like Ge, Sn and Pb. 
Indeed, in Figure \ref{Wcortes} we plot entropy for magnetic field
$B=10^{-2}$   (thick lines) and $B=10^{-3}$ T  (thin lines) and note that entropy slopes at $\Delta_z^{(0)}$ decrease 
with $B$ but they have a common critical entropy \eqref{maxent}. For electrons
(holes), minimum (maximum) entropies for a given Landau level $n$ are attained
at asymptotic values $\Delta_z\to\infty$,  and maximum (minimum)
entropies are attained at $\Delta_z\to -\infty$. The minimum and maximum entropies turn out to 
depend only on $|n|$ as
\begin{eqnarray}
W^{2,s}_{n,\mathrm{min}}&=&
-\ln\left(\frac{|n|\Gamma(|n|-\frac{1}{2})}{2\sqrt{\pi}\Gamma(|n|+1)}\right)\nonumber\\
W^{2,s}_{n,\mathrm{max}}&=&
-\ln\left(\frac{(|n|-\frac{1}{2})\Gamma(|n|-\frac{1}{2})}{2\sqrt{\pi}\Gamma(|n|+1)}\right),\label{maxminent}
\end{eqnarray}
where $\Gamma$ denotes the usual gamma function. Note that $W^{2,s}_{n+1,\mathrm{min}}=W^{2,s}_{n,\mathrm{max}}$, as can also be seen in Figure \ref{Wcortes}. 
For example, $W^{2,s}_{\pm 1,\mathrm{min}}=\ln(2)$ (which means that the  Hamiltonian eigenstate 
$|\pm 1\rangle_{s}$  is a coherent state of electrons for $\Delta_z\to\infty$ and a coherent state of holes for $\Delta_z\to-\infty$) 
and  $W^{2,s}_{\pm 1,\mathrm{max}}=2\ln(2)$ (which means that 
$|\pm 1\rangle_{s}$  is a ``cat state'' of electrons for $\Delta_z\to-\infty$ and a cat state of holes for $\Delta_z\to\infty$). 
Indeed, as commented after \eqref{moment2n}, the minimum entropy  of $W^2_{\mathrm{min}}=\ln(2)$ is attained for coherent states. 
The so-called ``cat states''  arise in many models undergoing a QPT (see e.g., \cite{husidi,husivi,LMG})  and are a superposition 
of two coherent (semiclassical) states with negligible overlap, so that their entropy is twice the minimum entropy, that is, $2\ln(2)$.

In Figure \ref{Renyi1} we plot the case $n=\pm 1$ for higher magnetic fields
($B=0.1$  and $B=1$ T in panel a) and the combined entropy of electrons plus holes, 
$W^{2,s}_n+W^{2,s}_{-n}$ (in panel b), which exhibits a maximum of $2C_n$ 
at the CNPs. For $n=\pm 1$, 
the maximum entropy of the electron-hole compound is $2C_1=2\ln(16/5)\simeq 2.3263$ (see Figure \ref{Renyi1}, panel b). The asymptotic behavior 
of the maximum combined entropy for large $n$ is $2C_n\simeq \ln(4\pi)+\ln(|n|)+O(1/|n|)^{3/2}$, which shows an increasing logarithmic 
behavior with $|n|$.

\begin{figure}
\includegraphics[width=8cm]{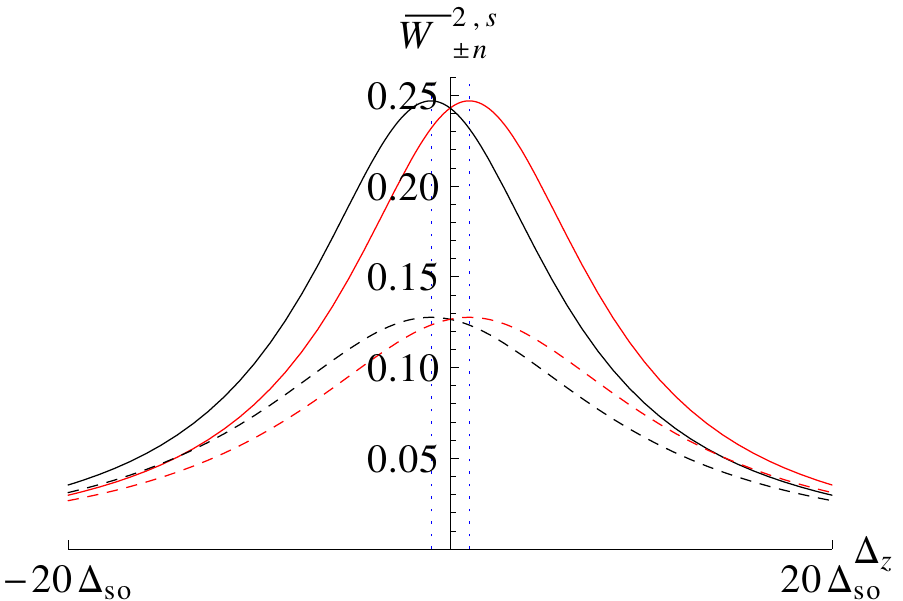}
\caption{ \label{Renyi3} Renormalized R\'enyi-Wehrl entropies \eqref{renorment} of the electron-hole compound  
for a Hamiltonian eigenstate $|n\rangle_{\xi,s}$ 
as a function of the electric potential $\Delta_z$ ($B=1$ T) for the Landau levels $n=1$ (solid lines) and $n=2$ (dashed lines),  
and valley $\xi=1$. 
Black and red curves correspond to spin $s=-1$ and $s=1$, respectively. Vertical blue lines indicate the CNPs.}
\end{figure}

The existence of a maximum electron plus hole entropy means that the combined wave function spreads across the phase space, 
taking up a bigger volume (it is more delocalized) at the CNP. In Figure \ref{Renyi3} we also plot 
the combined renormalized entropy 
\begin{equation}
 \overline{W}^{2,s}_{\pm n}=W^{2,s}_n+W^{2,s}_{-n}-W^{2,s}_{n,\mathrm{min}}-W^{2,s}_{n,\mathrm{max}}\label{renorment}
\end{equation}
of particles plus holes for $n=1$ (solid lines) and 
$n=2$ (dashed lines) together. We see that entropy ``hats'' become flatter and flatter as $n$ increases.

The global behavior of the entropies $W_n^{2,s}$, as a function of the
  electric field strength, exhibits some common features with the band inversion behavior for the $n=0$ 
  Landau level (see Fig. \ref{energias} and comments at the end of Section \ref{Hamilsec}). 
  Indeed, defining the renormalized entropy 
\begin{equation}
\widetilde{W}_n^{2,s}(\Delta_z)={W}_n^{2,s}(\Delta_z)-C_n=-
\ln\left(\frac{{M}_n^{2,s}(\Delta_z)}{{M}_n^{2,s}(\Delta_z^{(0)})}\right), 
\end{equation}
one can see that the quantities $\widetilde{W}_n^{2,s}(\Delta_z)$ and $\widetilde{W}_n^{2,-s}(\Delta_z)$ 
(for $n=\pm 1,\pm 2,\dots$) have the same sign in the BI phase
($|\Delta_z|>\Delta_\mathrm{so}$), and different 
sign in the TI phase ($|\Delta_z|<\Delta_\mathrm{so}$). See e.g. Figure \ref{Renyi1} for $B=1$ T (thick lines). 
Therefore, this ``entropy-inversion'' provides a clear characterization of each phase.  

Additionally, taking into account that $W_n^{2,s}$ measures the area occupied by the eigenstate $|n\rangle_s$ in phase space, we can say 
that, in the BI phase, an spreading (resp. localization) effect can be observed in the electrons
(resp. holes) for $\Delta_z<-\Delta_\mathrm{so}$ and in the holes (resp. electrons) for
$\Delta_z>\Delta_\mathrm{so}$;  in the TI phase there is a spreading effect in
the electrons with spin up and holes with spin down, and a localization in electrons
with spin down and holes with spin up. We observe that the lower the magnetic field
strength, the sharper these spreading and localization effects are.

\section{Conclusions}

In this work we have shown that the Husimi function is an appropriate tool to study  
topological-band insulator transitions in silicene and other gapped
isoestrucural Dirac materials with intrinsic spin-orbit coupling as germanene, stantene or Pb.
In particular, the R\'enyi-Wehrl entropy of the Husimi function of order two (related to the inverse participation
ratio) has been used to analyze the TI to BI quantum phase transition, 
showing the critical points  of this transition clearly as  a crossing/inversion 
behavior of the electron and hole entropies at the charge neutrality point. Critical \eqref{maxent} and asymptotic \eqref{maxminent} values 
of electron-hole entropies only depend on the Landau level $n$, and not on any other physical quantities like: 
magnetic strength, Fermi velocity, electron charge, etc. Therefore, these entropic measures provide a universal 
characterization of the topological-band insulator QPT, which is shared with other two-dimensional crystals 
like Ge, Sn and Pb. 

We want to emphasize that all the descriptions in terms of R\'enyi-Wehrl entropies that we have
presented in this work  will be valid in (isoestructural) 2D gapped Dirac materials
with a strong intrinsic spin-orbit  interaction. Actually, all the
characterizations studied here will apply to Ge, Sn and Pb
counterparts. In fact, all the Figures  will be valid by simply replacing the
value of $\Delta_\mathrm{so}$ in each case: $\Delta_\mathrm{so}^\mathrm{Ge}=11.8$ meV for
germanene, $\Delta_\mathrm{so}^\mathrm{Sn}=36.0$ meV for stanene,
$\Delta_\mathrm{so}^\mathrm{Pb}=207.3$ meV for Pb \cite{nature}. Moreover, Dirac equation (and modifications) provides 
an effective model for a large family of topological insulators, and an entropic description like the one 
done in this letter could also sheed new light on the better understanding of topological phases in these 
physical systems. This deserves a separated and more thorough study and it will be 
the subject of future work.

\section*{Acknowledgments}
  The work was supported by 
the Spanish Projects MICINN
FIS2011-24149, CEIBIOTIC-UGR PV8, the Junta de Andaluc\'{\i}a projects FQM.1861 and FQM-381.

\end{document}